\documentclass[aps,prl,twocolumn,superscriptaddress,showpacs]{revtex4}
\usepackage{graphicx}
\begin{document}


\title{Field-Induced Magnetic Order and Simultaneous Lattice Deformation in TlCuCl$_3$}
\author{O.~Vyaselev}
\affiliation{Institute for Solid State Physics, University of Tokyo, Kashiwanoha, Kashiwa, Chiba 277-8581, Japan}
\affiliation{Institute of Solid State Physics, Russian Academy of Science, Chernogolovka Moscow District 142432, Russia}
\author{M.~Takigawa}
\email[]{masashi@issp.u-tokyo.ac.jp}
\affiliation{Institute for Solid State Physics, University of Tokyo, Kashiwanoha, Kashiwa, Chiba 277-8581, Japan}
\author{A.~Vasiliev}
\affiliation{Department of Low-Temperature Physics, Moscow State University, Moscow 119899, Russia}
\author{A.~Oosawa}
\author{H.~Tanaka}
\affiliation{Department of Physics, Tokyo Institute of Technology, Oh-okayama, Meguro-ku, Tokyo 152-8551, Japan}

\date{\today}

\begin{abstract}
We report the results of Cu and Cl nuclear magnetic resonance experiments (NMR) and
thermal expansion measurements in magnetic fields in the coupled dimer spin system TlCuCl$_3$.
We found that the field-induced antiferromagnetic transition as confirmed by the splitting of NMR lines
is slightly discontinuous.  The abrupt change of the electric field gradient at the Cl sites,
as well as the sizable change of the lattice constants, across the phase boundary indicate that
the magnetic order is accompanied by simultaneous lattice deformation.
\end{abstract}


\maketitle

Quantum phase transitions in spin systems induced by magnetic field have
attracted considerable recent interest~\cite{rice031,chaboussant981,kodama021,shimamura971,
oosawa991,giamarchi991,nikuni001,oosawa011}.  An important class of materials is formed by antiferromagnets
with a collective singlet ground state at zero field separated by a energy gap $\Delta$ from
the first excited triplet.  A magnetic field $H$ reduces the gap as
$\Delta(H) = \Delta - g\mu_{B}H$.  At $T=0$, a finite magnetization appears above the critical field
$H_c=\Delta/g\mu_B$.  Since antiferromagnetic interactions acting among the field-induced
spin moments will tilt the magnetization into two sublattices,
a N\'{e}el order perpendicular to the external field is expected above $H_c$.
This process can be described as the Bose-Einstein condensation of triplet
magnons~\cite{giamarchi991,nikuni001}, since the creation operator of the lowest energy magnon is
equivalent to the transverse component of the staggered magnetization.
With further increasing the field, some materials show magnetization plateaus at fractional values of the fully saturated
moment~\cite{shimamura971,kodama021}, indicating localization of triplets and formation of a superstructure.
A wider variety of exotic phases may be expected in the presence of spin-phonon coupling or orbital degeneracy.
It is important that these phase transitions can be controlled precisely
by tuning the field value, which sets the chemical potential for triplets.  Thus
spin systems in magnetic field provide valuable opportunities to test theories on strongly correlated
quantum systems.

\begin{figure}[b]
\includegraphics*[scale=0.3]{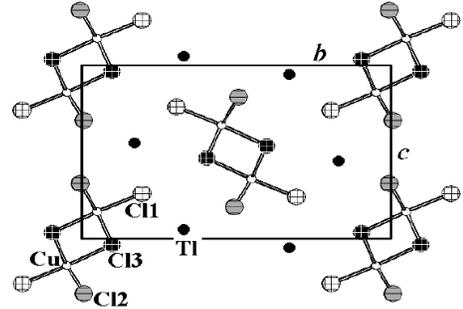}
\caption{\label{structure} The crystal structure of TlCuCl$_3$ viewed along the $a$-axis.}
\end{figure}
TlCuCl$_3$ is a system of three-dimensionally (3D) coupled dimers formed by $S$=1/2 spins of Cu$^{2+}$ ions
with a singlet ground state~\cite{oosawa991,oosawa011}.  Although the crystal structure
(monoclinic $P2_1/c$ space group) contains Cu$_2$Cl$_6$ zigzag chains extending along the
$a$-axis as a subunit (Fig.~\ref{structure})~\cite{willett631}, inelastic neutron scattering experiments
revealed fully three-dimensional dispersions of triplet magnons~\cite{oosawa021,cavadini011},
from which the exchange interactions have been determined.  The minimum excitation energy
$\Delta \simeq 0.5$~meV occurs at $Q=(0,0,1)$.  The static transverse magnetic order has been observed
at the same $Q$ above a temperature-dependent critical field $H_c(T)$ with $(g/2)H_c(0)=5.7$~T~\cite{tanaka011}.  Since
the dispersion width (7~meV) is much larger than the excitation gap at zero field,
the density of triplets is small when the field is close to $H_c$ and the system can be considered
as interacting dilute Bose gas.  In fact, mean field~\cite{nikuni001} and
RPA analysis~\cite{matsumoto021} of boson model were applied
successfully to explain the peculiar shape of the magnetization curve, as well as the
excitation spectrum as a function of magnetic field~\cite{cavadini021,ruegg031,ruegg032} in a semi-quantitative manner.

In this letter we report the results of Cu and Cl nuclear magnetic resonance (NMR) experiments and
thermal expansion measurements on TlCuCl$_3$ near the critical field.  The antiferromagnetic
order was confirmed by the splitting of NMR lines.  The staggered magnetization develops 
discontinuously and the paramagnetic and
antiferromagnetic phases coexist in a narrow region in the $H-T$ plane, pointing to a weakly first order
phase transition. We found also that the electric field gradient (EFG) at the Cl sites changes abruptly across the phase boundary,
indicating that a lattice deformation occurs simultaneously at the magnetic transition.  This is further confirmed by a sizable change of
the lattice parameter measured by strain gauge technique in magnetic field .

As shown in Fig.~\ref{structure}, a monoclinic unit-cell of TlCuCl$_3$ contains four formula units.   When the magnetic field
is along or perpendicular to the $b$-axis, the four units give identical NMR lines.
All the measurements were performed in the external field perpendicular both to the $b$-axis and
approximately to the (10$\bar 2$) plane.  Thus one Cu and three Cl sites should be distinguished.
There are six Cu NMR lines in the absence of magnetic order: each of the two spin-3/2 isotopes, 
$^{63}$Cu and $^{65}$Cu, generates three lines split by electric quadrupole interaction.
The Cu NMR data were taken on the central transition (--1/2$\leftrightarrow$1/2) line from $^{63}$Cu.
For Cl NMR, we found that 18 lines (three quadrupole split lines for each of two spin-3/2 isotopes, $^{35}$Cl and
$^{37}$Cl, from three sites) are congested in a relatively narrow frequency range.  Fortunately, a pair of outermost
lines corresponding to the low-frequency quadrupole satellite of $^{37}$Cl and high-frequency satellite
of $^{35}$Cl from the same Cl site are isolated from other lines.  We used these lines to obtain Cl NMR
data, although we have been unable to assign the specific sites to which these lines correspond.
The thermal expansion was measured by the strain gauge technique.  The strain gauge was calibrated using
the spontaneous strain along the $a$-axis at the phase transition in $\alpha'$-NaV$_2$O$_5$~\cite{koeppen981}.
The gauge was attached to the (10$\bar 2$) plane along the $b$-axis. To correct for the magneto-resistance
of the gauge, the curves taken in different fields have been shifted to merge at high temperatures.

\begin{figure}[t]
\includegraphics*[scale=0.4]{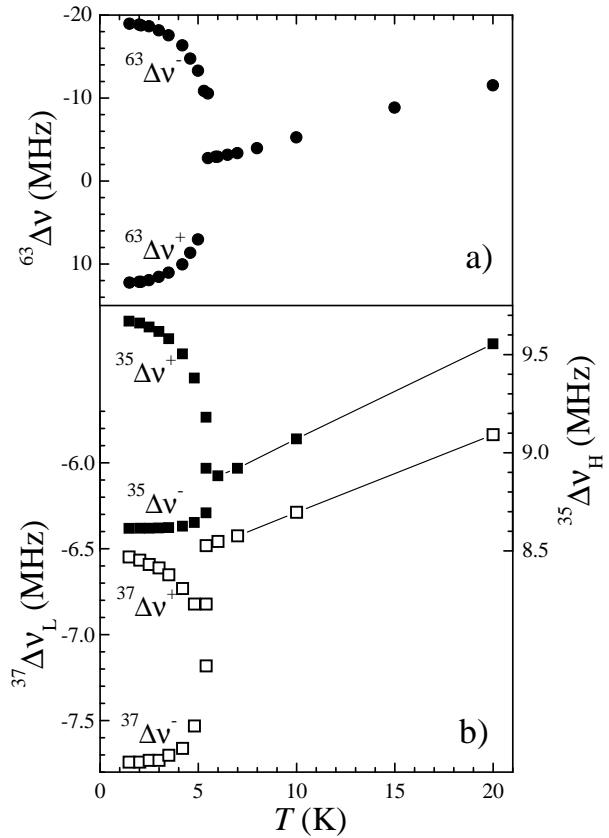}%
\caption{\label{RawShift} Temperature dependencies of NMR frequency shifts at $H = 8.56$~T. (a):
Central transition of $^{63}$Cu. (b): High-frequency satellite of $^{35}$Cl (closed squares,
right axis) and low-frequency satellite of $^{37}$Cl (open squares, left axis).}
\end{figure}

Figure~\ref{RawShift} shows the frequency shifts for the central transition of $^{63}$Cu (a) and for the pair of Cl
satellites (b) as a function of temperature at $H = 8.56$~T.  The shifts are defined as
${^{n}\Delta}\nu= {^{n}\nu} - {^{n}\gamma} H$, where $^{n}\nu$ is the peak frequency of $n$-th nuclei with
the gyromagnetic ratio $^{n}\gamma$.  The main feature is the splitting of all lines into two branches ${^{n}\Delta}\nu^\pm$
below 5.6~K, which coincides the N\'{e}el temperature ($T_{\rm{N}}$) for $H\parallel [10\bar{2}]$ at 8.56~T \cite{oosawa011}.

Generally, the shift of $^{63}$Cu NMR is the sum of the magnetic
and quadrupolar parts. For the central transition, the quadrupolar
part contains only the second and higher order terms
${^{63}\Delta}\nu_{\rm Q}^{(2)}$.  The magnetic part is due only to
the uniform magnetization ${^{63}\Delta}\nu_{\rm{u}}$ above
$T_{\rm{N}}$, but contribution from the staggered magnetization
${^{63}\Delta}\nu_{\rm{s}}$ adds below $T_{\rm{N}}$. Thus
${^{63}\Delta}\nu={^{63}\Delta}\nu_{\rm{u}} + {^{63}\Delta}\nu_{\rm Q}^{(2)}$ above $T_N$ and
${^{63}\Delta}\nu^\pm={^{63}\Delta}\nu_{\rm{u}} + {^{63}\Delta}\nu_{\rm Q}^{(2)} \pm {^{63}\Delta}\nu_{\rm{s}}$ below
$T_N$. It should be noted that even though the staggered
magnetization is perpendicular to the external field, anisotropic
hyperfine interactions generally produce a finite component of
staggered local field at the nuclei, $^{n}h_{\rm{s}} = {^{n}\Delta}\nu_{\rm{s}}/{^{n}\gamma}$, parallel to the external
field. The staggered contribution below T$_N$ cancels for the mean value
$\overline{{^{63}\Delta}\nu}\equiv({^{63}\Delta}\nu^++{^{63}\Delta}\nu^-)/2=
{^{63}\Delta}\nu_{\rm{u}}+{^{63}\Delta}\nu_{\rm{Q}}^{(2)}$.

\begin{figure}[b]
\includegraphics*[scale=0.3]{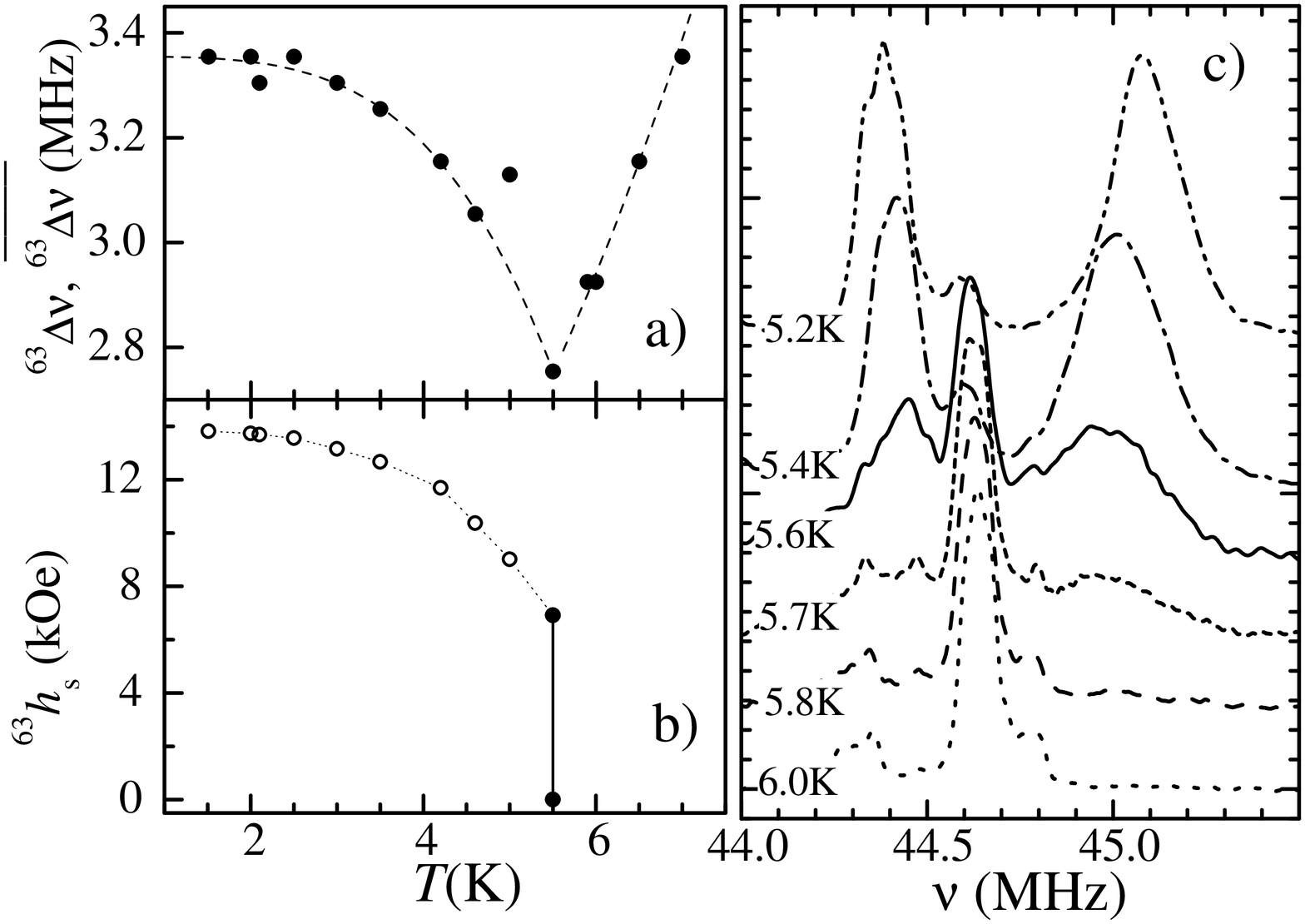}
\caption{\label{SpecStgUnif} (a) Uniform frequency shift for the $^{63}$Cu central transition. The line is a guide to the eye.
(b) The staggered hyperfine field obtained from the splitting of the $^{63}$Cu central line. At 5.5~K, the spectrum consists
of unsplit ($^{63}h_{\rm{s}}=0$) and split (finite $^{63}h_{\rm{s}}$) lines.  (c)  Evolution of the high-frequency satellite $^{35}$Cl
NMR spectrum across the phase boundary.  All data were taken at $H = 8.56$~T.}
\end{figure}
We plot ${^{63}\Delta}\nu$ (for $T \geq T_N$), $\overline{^{63}\Delta\nu}$ (for $T \leq T_N$)
and $^{63}h_{\rm{s}}$ against temperature in Fig.~\ref{SpecStgUnif} (a) and (b).
Since the temperature variation of ${^{63}\Delta}\nu_{\rm Q}^{(2)}$ below 7~K is expected to be
negligibly small, the change of ${^{63}\Delta}\nu$ and $\overline{^{63}\Delta\nu}$
should be due to the temperature variation of the uniform magnetic shift, ${^{63}\Delta}\nu_{\rm{u}}$.
Indeed, the plot in Fig.~\ref{SpecStgUnif}(a) is quite similar to the temperature dependence of the
bulk magnetization~\cite{oosawa991}, including the characteristic cusp-like minimum at $T_{\rm{N}}$.

The staggered local field $^{63}h_{\rm{s}}$ rises very sharply at $T_{\rm{N}}$. The
spectrum at 5.5~K consists of a single line originating from the paramagnetic
phase and a pair of split lines coming from the antiferromagnetic phase.
With increasing temperature, the intensity of the latter vanishes
within about 0.2~K. The same feature was seen in the evolution of
the Cl NMR spectra as shown in Fig.~\ref{SpecStgUnif}(c).  At
5.6~K, one can clearly see the coexistence of a single peak and a
pair of split peaks.  Moreover, tracking the temperature evolution
from 5.2~K upwards, one notices that the split lines do not merge
to become a single line. Instead, the intensity of the split lines
vanishes between 5.7 and 5.8~K, while the interval of the
splitting is still finite. We observed no hysteresis for up and
down temperature sweep.  The coexistence of two phases is clearly
due to some kind of disorder.  The appearance of the split
spectrum, however, indicates that the staggered magnetization
develops discontinuously upon entering into the antiferromagnetic
phase.  Thus the transition is weakly first order.  We also
observed the same feature at a fixed temperature $T = 3.5$~K with
increasing magnetic field across the phase boundary near 6.3~T.

We emphasize that the discontinuity of the transition observed by NMR is not in
contradiction with the continuous development of the antiferromagnetic Bragg peaks observed by neutron diffraction
experiments~\cite{tanaka011}.  The intensity of the Bragg peaks 
represents square of the staggered moment averaged over the entire sample volume.  If the volume fraction
of the antiferromagnetic phase changes continuously over a finite temperature window near $T_{\rm{N}}$ as
evidenced by the evolution of the Cl NMR spectrum, a locally discontinuous transition should look like a
continuous one when averaged over the entire volume.

\begin{figure}[t]
\includegraphics[scale=0.35]{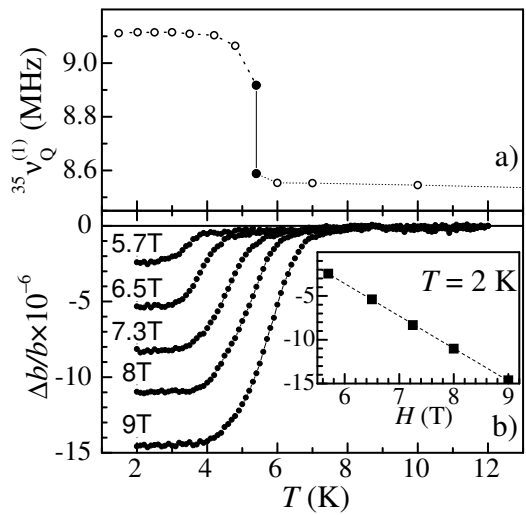}%
\caption{\label{Fig4} (a) First-order quadrupole shift for $^{35}$Cl extracted with
eq.~(\ref{QShift}) from the data in Fig.~\ref{RawShift}(b). (b) $b$-axis thermal
expansion in different magnetic fields. The insert shows the $b$-axis strain
\textit{versus} magnetic field at 2~K.}
\end{figure}

Another important feature of the field-induced transition was revealed by the behavior of the quadrupole splitting
of the Cl NMR spectrum.  Above $T_{\rm{N}}$, the shifts for the high-frequency $^{35}$Cl satellite and 
low-frequency $^{37}$Cl satellites plotted in Fig.~\ref{RawShift}b are expressed by
\begin{eqnarray}
{^{35}\Delta}\nu_{H}&=&{^{35}\Delta}\nu_{\rm{u}}+{^{35}\Delta}\nu_{Q}^{(2)}+{^{35}\Delta}\nu_{Q}^{(1)},\nonumber\\
{^{37}\Delta}\nu_{L}&=&{^{37}\Delta}\nu_{\rm{u}}+{^{37}\Delta}\nu_{Q}^{(2)}-{^{37}\Delta}\nu_{Q}^{(1)},
\label{Sat}
\end{eqnarray}
\noindent where $\Delta\nu_Q^{(1)}$ is the first-order quadrupole shift. This is determined
by the magnitude of the electric field gradient along the external field direction.  The temperature dependence
of $\Delta\nu_Q^{(1)}$ reflects the change of lattice parameters and ionic charge distribution within a unit cell.
Below $T_{\rm{N}}$, the staggered shift adds to eqs.~(\ref{Sat}): ${^{35}\Delta}\nu_H^\pm={^{35}\Delta}\nu_H\pm{^{35}\Delta}\nu_{\rm{s}}$,
${^{37}\Delta}\nu_L^\pm={^{37}\Delta}\nu_L\pm{^{37}\Delta}\nu_{\rm{s}}$. Note that $\Delta\nu_{\rm{s}}$ cancels for the mean values,
$\overline{^{n}\Delta\nu_{H(L)}}\equiv({^{n}\Delta}\nu_{H(L)}^++{^{n}\Delta}\nu_{H(L)}^-)/2$,
so below $T_{\rm{N}}$ $\overline{{^n\Delta}\nu_{H(L)}}$ replaces the left-hand side of
eqs.~(\ref{Sat}).  Using the isotopic ratio of each term in eqs.~(\ref{Sat}) 
\begin{eqnarray}
{^{37}\Delta}\nu_{\rm{u}}/{^{35}\Delta}\nu_{\rm{u}}&=&{^{37}\gamma}/ {^{35}\gamma}\equiv r_\gamma\simeq0.83,\nonumber\\
{^{37}\Delta}\nu_{Q}^{(1)}/{^{35}\Delta}\nu_{Q}^{(1)}&=&{^{37}Q}/ {^{35}Q}\equiv r_Q\simeq0.79,\label{Scaling}\\
{^{37}\Delta}\nu_{Q}^{(2)}/{^{35}\Delta}\nu_{Q}^{(2)}&=&r_Q^2/r_\gamma\simeq0.87,\nonumber
\label{ratio}
\end{eqnarray}
\noindent where $^nQ$ is the quadrupole moment of the $n$-th nucleus,  
one obtains ${^{35}\Delta}\nu_{Q}^{(1)}$ as   
\begin{eqnarray}
\left( r_{\gamma}{^{35}\Delta}\nu_H - {^{37}\Delta}\nu_L \right) / \left( r_{\gamma}+r_{Q} \right) \nonumber\\
= {^{35}\Delta}\nu_{Q}^{(1)} + {^{35}\Delta}\nu_{Q}^{(2)} \left( r_{\gamma}-r_{Q} \right)/r_{\gamma} \approx {^{35}\Delta}\nu_{Q}^{(1)} 
\label{QShift}
\end{eqnarray}
\noindent for $T>T_{\rm{N}}$.  For $T<T_{\rm{N}}$, ${^{n}\Delta}\nu_{H(L)}$ should be replaced
by $\overline{{^{n}\Delta}\nu_{H(L)}}$. Since ${^{35}\Delta}\nu_{Q}^{(2)} \ll {^{35}\Delta}\nu_{Q}^{(1)}$ 
and $\left( r_{\gamma}-r_{Q} \right)/r_{\gamma}=0.05$, the second term in the right-hand side 
can be safely neglected.   

The temperature dependence of ${^{35}\Delta}\nu_{Q}^{(1)}$ is plotted in Fig.~\ref{Fig4}(a). 
There is a clear jump in ${^{35}\Delta}\nu_Q^{(1)}$
at $T_N$, indicating a sudden change in the atomic positions and/or lattice parameters.
The change of lattice parameters was directly detected by the thermal expansion measurements.
In Fig.~\ref{Fig4}(b), the relative change of the crystal size along the $b$-axis, $\Delta b/b$, is
plotted against temperature for different magnetic fields.  Anomalous expansion is observed below a field-dependent
onset temperature, which coincides with $T_N$.  The magnitude of the strain is linear in field as shown in the insert in
Fig.~\ref{Fig4}(b).  Measurement of the strain along the $a$-axis has shown that $\Delta a/a$ is negative and
an order of magnitude smaller than $\Delta b/b$.  The $c$-axis strain could not be measured properly.
The change of the lattice parameters, however, is very small and this alone would not be sufficient to account for the
large change of ${^{35}\Delta}\nu_{Q}^{(1)}$.  There must be sizable displacement of the ionic positions
within a unit cell.

The results presented so far convincingly demonstrate that lattice deformation occurs
simultaneously with the magnetic phase transition.
Thus the Bose-Einstein condensation scenario within a pure spin model is not adequate and spin-phonon
coupling has to be taken into account for quantitative description of the field induced
phase transition. Strong spin-phonon coupling has been also suggested by the recent
observation of pressure-induced antiferromagnetic order~\cite{oosawa031}.  
It is known in classical magnets that strong enough magneto-elastic coupling turns an otherwise 
second order magnetic transition into a first order one accompaning magnetostriction~\cite{ito781}.  
It is likely that similar mechanism works in the present case of a quantum phase transitions. 
However, It is still possibile that specifc interactions among bosons in this material is the 
driving force for the first order transition. 

The fact that the transition is only marginally discontinuous is also indicated by quasi-critical
slowing down of the spin fluctuations observed by the nuclear spin-lattice relaxation rate
measurements.
\begin{figure}[t]
\includegraphics[scale=0.25]{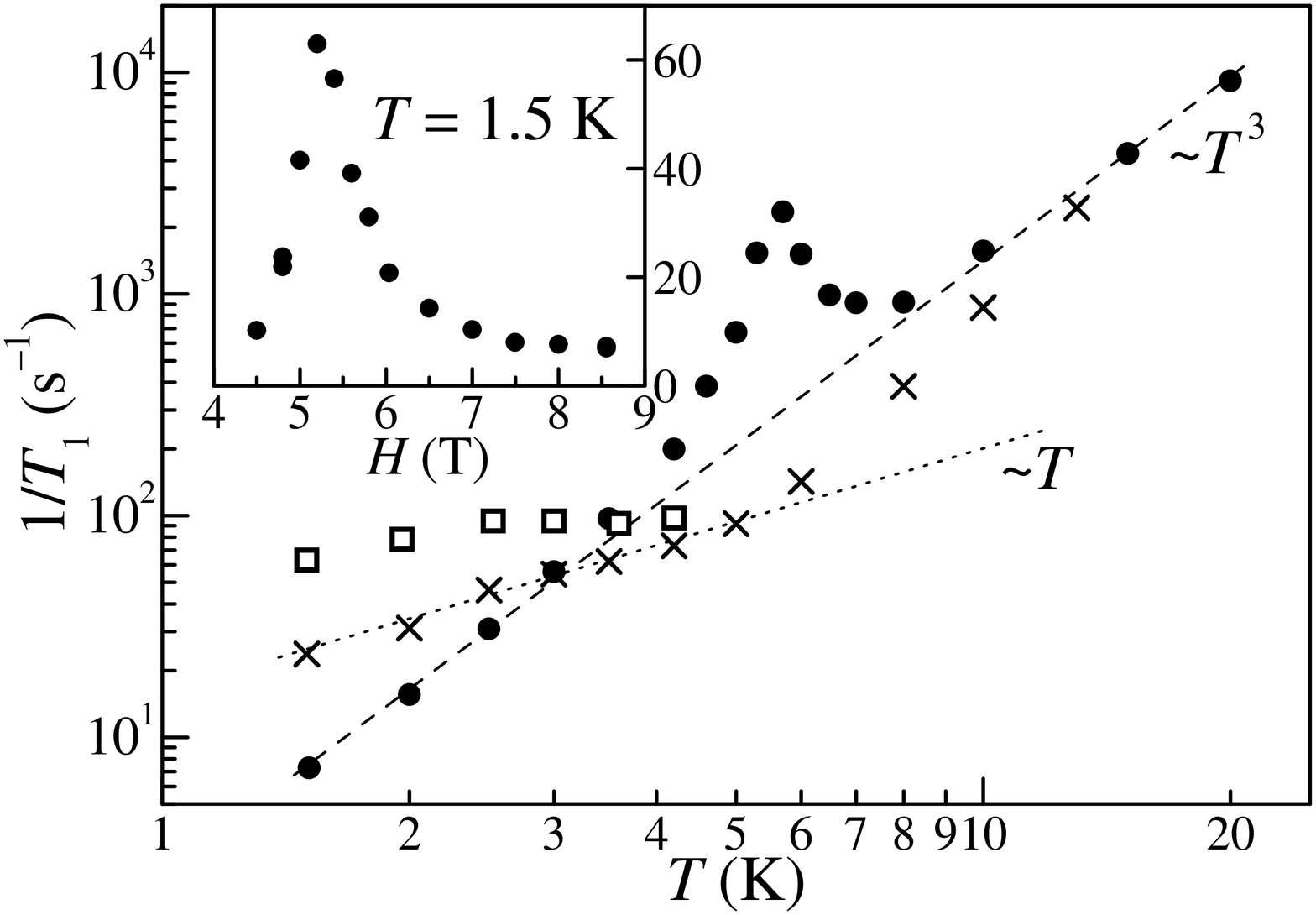}
\caption{\label{FigT1}Temperature dependence of $^{63}$Cu spin-lattice relaxation rate at $H$ = 5.2~T
(squares), 4.8~T (crosses), and 8.5~T (circles). Insert: Field dependence of the relaxation rate at 1.5~K.}
\end{figure}
The spin-lattice relaxation rate, $1/T_1$, of $^{63}$Cu nuclei is plotted against temperature in Fig.~\ref{FigT1} 
for the external fields of 4.8~T, 5.2~T (the zero-temperature critical field), and 8.5~T. 
The inset shows the field dependence of $1/T_1$ at 1.5~K.
The data at 8.5~T clearly shows a sharp peak at $T_N$.
Above 8~K, $1/T_1$ increases approximately as $T^3$.  Such temperature
dependence indicate that the major weight of the spin fluctuations has energy scale much larger than
$T_N$.  At lower temperatures, the relaxation rate at 4.8~T switches to linear $T$-dependence, 
and $1/T_1$ at 5.2~T is nearly independent of temperature. The relaxation rate at the phase boundary is sharply peaked, as
seen in the $T$-dependence at $H = 8.5$~T and in the $H$-dependence at $T = 1.5$~K.  Such behavior of $1/T_1$ is
due to slowing down of the fluctuations of staggered moments in the vicinity of the phase boundary, which is
common for continuous antiferromagnetic transitions.

In conclusion, we clarified some new features of the field-induced magnetic order in TlCuCl$_3$.
Splitting of the NMR peaks in the field-induced phase reveals a long range N\'{e}el order. 
The staggered local field is shown to arise discontinuously, indicating that the transition is weakly first-order.  Sizable
deformation of the lattice due to the magnetic ordering is detected in both NMR and thermal expansion
measurements, which implies strong spin-phonon coupling and possibly explains the first-order nature of the
transition. The spin-lattice relaxation rate is sharply peaked at the phase boundary due to quasi-critical slowing down
of the fluctuations of staggered magnetization.

\begin{acknowledgments}
The authors acknowledge fruitful discussions with K.~Ueda and M.~Zhitomirsky. The work is supported by
the Grant-in-Aid for Scientific Research on Priority Area (B) on ``Field-induced new quantum phenomena in magnetic
systems" from Ministry of Education, Culture, Sports, Science, and Technology of Japan. The research activity of
O.V. in Japan was supported by the Japan Society for Promotion of Science.
\end{acknowledgments}


\end{document}